# Photoionization of temperature-controlled nanoparticles in a beam: Accurate and efficient determination of ionization energies and work functions

Atef A. Sheekhoon,[1,a)] Abdelrahman O. Haridy,[1,a)] Sebastian Pedalino,[2] Vitaly V. Kresin[1,b)]

[1]Department of Physics and Astronomy, University of Southern California,
Los Angeles, 90089-0484, USA

[2]Faculty of Physics and Vienna Doctoral School in Physics, University of Vienna,
Boltzmanngasse 5, 1090 Vienna, Austria

A beam of free alkali metal nanoparticles is produced by a condensation source, passed through a thermalizing tube adjustable over a broad temperature range, and ionized by tunable light. High stability of the particle flux and an automated data acquisition routine allow efficient collection of photoionization yield curves. A careful fit of the data to the universal Fowler function makes it possible to obtain nanoparticle ionization energies, and from those the metal work functions, with ~0.2% precision. The experimental arrangement, nanoparticle thermalization rates, and ionization threshold analysis are described in detail. The use of ultrapure and temperature-controlled gas-phase nanoparticles facilitate the analysis of electronic properties, such as work functions, and of their interplay with thermal lattice dynamics.

a)These authors have contributed equally
b)Author to whom correspondence should be addressed: *kresin@usc.edu*

## I. INTRODUCTION

The minimum energy required to remove an electron from a metal surface (the work function, WF) or from a metal nanoparticle (the ionization energy, IE) is not only a keystone concept in quantum theory but also one of the principal characteristics of the electronic properties of a material. Not surprisingly, the list of experimental and theoretical approaches for the determination of work functions is vast. A concise but representative recent listing of such techniques can be found, for example, in Ref. 1, and an earlier review in Ref. 2.

Since WF measurements can be strongly affected by even minute amounts of surface contamination, accurate determination of this quantity is a challenge, especially for reactive materials. For example, the literature values of lithium metal WF vary by as much as 10%.[3] The use of free nanoparticles in a molecular beam allow one to circumvent this challenge: their flight time through the evacuated apparatus takes only milliseconds, which is short enough that contamination is easily avoided.

As described in this paper, a careful implementation of this method enables a determination of alkali-metal IEs and WFs with high reproducibility, with a precision of better than 0.2% (several meV), and covering a wide range of material temperatures. The measurement makes use of photoionization spectroscopy in combination with a stable source of thermalized nanoparticles, sensitive ion detection, and systematic ionization threshold analysis. Thanks to the attained high precision and reproducibility, it becomes possible to investigate subtle but informative interplay between ionization thresholds and lattice dynamics. This can include, for example, the influence of thermal expansion and structural phase transitions on the WF.

This paper is organized as follows. Section II describes the experimental arrangement and discusses aspects of nanoparticle thermalization and size determination. Section III describes the data analysis procedure, including threshold effects and finite size corrections. Section IV presents a set of work functions deduced from the data, and Section V contains the conclusions.

## II. DESIGN OF THE EXPERIMENT

### A. Outline of the measurement

The photoionization measurements is outlined in Fig. 1. (Individual components will be described in more detail below.) Metal nanoparticles are produced using a vapor aggregation source. They are carried by a flow of helium gas through a thermalization tube and into vacuum, forming a directed beam. The particles then traverse several vacuum chambers of a 2-meter-long nanocluster beam apparatus and enter the photoionization and detection chamber. Here they are ionized by an arc lamp through a monochromator, and the resulting positive ions are detected by a Daly dynode-photomultiplier ion-counting arrangement.[4] The yield of ions is equivalent to that of photoelectrons, hence the near-threshold behavior of both is identical.

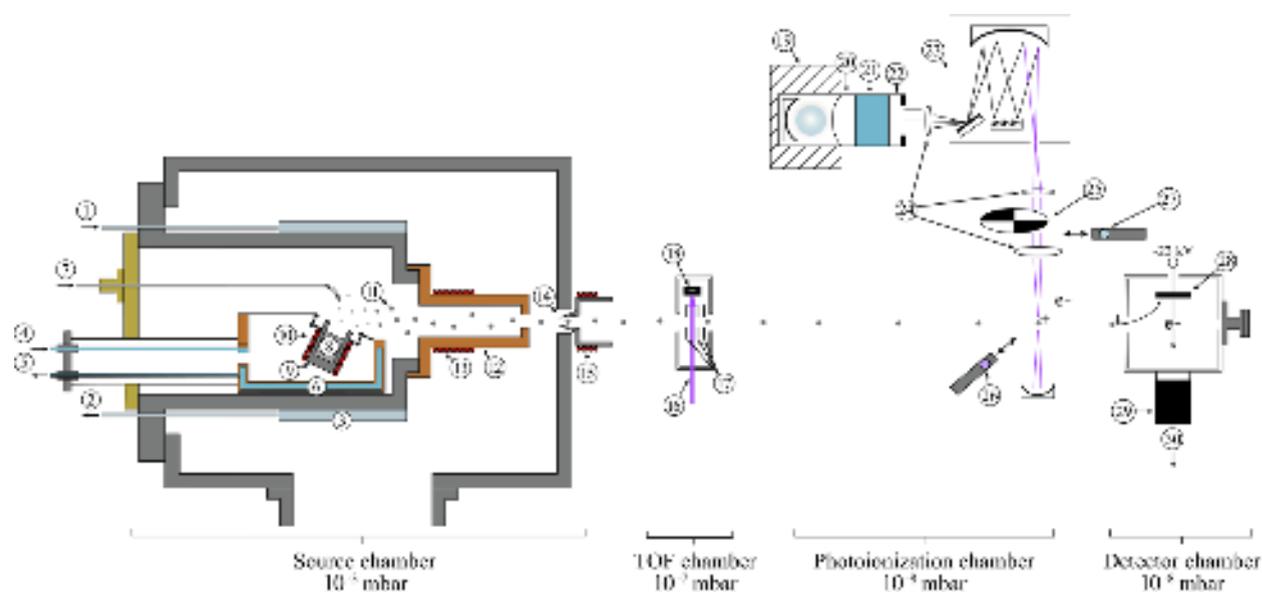

**FIG. 1**. An outline of the photoionization setup (not to scale). 1,2: Liquid nitrogen inlet and outlet; 3: liquid nitrogen jacket; 4,5: water inlet and outlet; 6: water-cooled box; 7: helium gas inlet; 8: alkali metal in the crucible; 9: crucible; 10: crucible heater; 11: vapor-condensed nanoparticles; 12: thermalization tube; 13: thermalization tube heater; 14: skimmer (3 mm opening); 15: skimmer heater; 16: Pulsed Nd:YAG laser for TOF ionization; 17: TOF plates; 18: laser energy meter; 19: UV arc lamp; 20: cylindrical lens; 21: IR water filter; 22: iris diaphragm; 23: monochromator with automated wavelength selection; 24: focusing lenses. 25: optical chopper; 26: photodiode on a motorized linear feedthrough; 27: UV LED on a motorized stage; 28: Daly dynode; 29: scintillator and photomultiplier tube; 30: signal pulses to computer.

Alternatively, the nanoparticles can be ionized by a pulsed laser in the extraction region past the skimmer, forming a collinear time-of-flight mass spectrometer used to calibrate their mass distribution. The average sizes in question comprised ~ 3000 – 10000 atoms, depending on the element and source conditions.

An arrangement of this type was previously used to study the alkali WFs over a limited set of temperatures[5,6]; the setup described here has been optimized toward significant improvements in source stability, temperature range and stabilization, and precision of IE determination.

### B. Nanoparticle source

Nanoparticle generation takes place in a gas aggregation source,[7] which involves evaporation of the material from a heated crucible followed by rapid condensation of the metal vapor within the flow of cold inert gas. The latter also transports nanoparticles out of the condensation chamber, terminating their growth.

In the present setup, the stainless steel crucible is open at the top and is mounted in a forward-tilted position inside a water-cooled copper box, see Fig. 2. Helium gas (5.0 grade purity) flow was directed toward the opening of the crucible. Its flow was set by a gas flow controller (Alicat MC-500SCCM-D) and set to 240 sccm. A constant supply of liquid nitrogen was fed into the outer jacket of the condensation chamber.[8] The oven body was electrically grounded to avoid occasional electrical discharges in the alkali vapor, likely caused by heater wire voltage leakage.

When using this source, the beam intensity remained exceptionally stable for at least 30 hours and decayed only when the metal load would start to run out.[9] This benefitted the precision of measurements by making it possible to avoid signal drift and to repeat each data point multiple times.

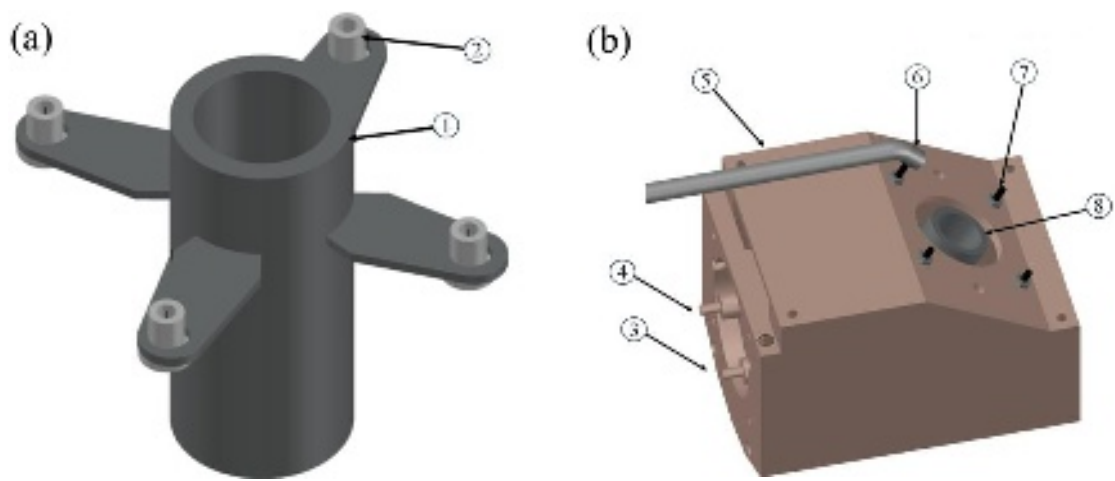

**FIG. 2.** The crucible (a) is mounted inside a water-cooled copper box (b) in a position inclined at 25° with respect to the beam direction. 1,8: crucible; 2: insulating Macor shoulder washers; 3,4: water inlet and outlet; 5: copper box; 6: helium gas inlet tube; 7: crucible mounting screws. The heating element is made of 24 AWG Kanthal wire wound in a coil and inserted into ceramic sleeves (High Temperature Nextel Braided ceramic and XS Silica sleeving, Omega Engineering) and wrapped around the crucible using high-temperature ceramic cement (Omegabond 600). Alternatively, the Kanthal wire can be threaded through 4-bore high-purity alumina tubing segments densely stacked around the circumference of the crucible. The crucible temperatures were maintained at 500°C for Li, 300°C for Na, and 200°C for K.

### C. Thermalization tube

A regulated thermalization tube is attached directly to the exit of the condensation chamber, presenting a smooth entryway to the nanoparticle/helium mixture. The tube is made of oxygen-free high-conductivity (OFHC) copper, has an overall length $l$ = 11 cm, and an inner diameter $d$ = 1.6 cm which narrows down to a 6 mm diameter, 1 cm long exit aperture. These dimensions were found to yield a suitable combination of beam intensity, gas flow and pressure, temperature uniformity, and thermalization conditions.[10]

In order to cover the full investigated temperature range, from 60 K to 390 K, the tubes' external construction was varied. For temperatures above 190 K, the tube is the same as that employed in Refs. 11,12. Its outside wall has a diameter of 3.8 cm and is wrapped with an electrical heater that is regulated by a temperature controller. To reduce thermal contact, a 6 mm thick Teflon spacer is placed between the base of the thermalization tube and the face of the condensation chamber. To enhance temperature stability, the tube is wrapped with superinsulation foil. Its temperature is monitored by three resistance temperature detectors (RTDs) spaced along its length, and is uniform to ±1 K.

In the low temperature tube design, the outer diameter is reduced to 2.5 cm, and thermal contact with the condensation chamber is further minimized by using a petal-shaped base and a patterned teflon spacer (Fig. 3). The tube is attached to the first stage of a closed cycle cryocooler (Leybold Coolpower 130 cold head with a COOLPAK 6000 MD compressor) by silver-coated copper braids. In addition, a heating element is wrapped around the base of the tube. Two silicon diodes (Lake Shore DT-670) are inserted into slots machined in the tube at opposite ends. One of the silicon diodes is connected to a cryogenic temperature controller (Lake Shore Model 340) which powers the heating element to set the tube temperature. The assembly is once again wrapped is superinsulation and is capable of reaching the aforementioned lowest temperature of 60 K, with a temperature variation of only ±1 K along the tube.

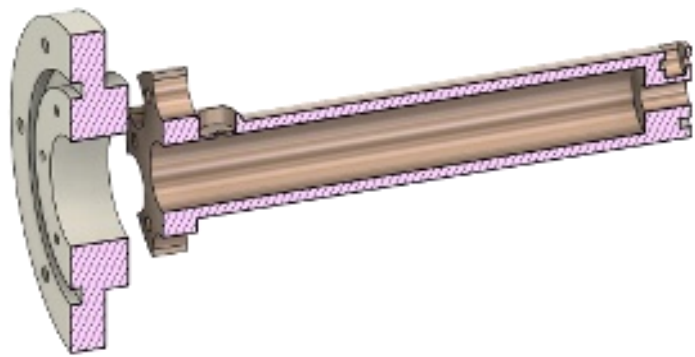

**FIG. 3**. A cross section of the low temperature thermalization tube. The two round wells in the tube wall are positions for the silicon diode tube temperature monitors. The groove on the base of the Teflon mounting flange (left) aims to minimize heat transfer between the source jacket and the thermalization tube.

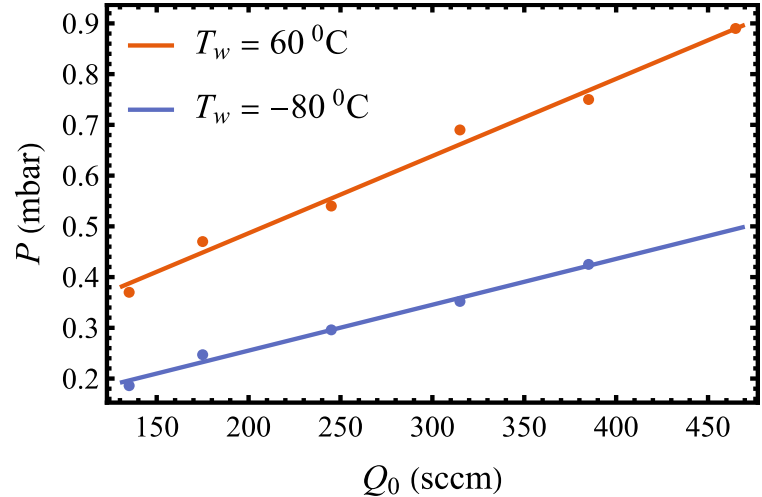

**FIG. 4**. Helium gas pressure $P$ inside the thermalization tube as a function of its flow rate $Q_0$ through the controller at two different wall temperatures $T_w$.

A critical question is whether the nanoparticles become reliably thermalized during their passage within the thermalization tube.

The tube walls interact with the flowing helium gas, and the latter then exchanges heat with the metal particles. By placing test temperature sensors along the center line of the tube axis while operating the source oven, we have directly verified that the helium gas equilibrates with the wall temperature $T_w$ to within a couple of degrees at most 4 cm (i.e., $\lesssim \frac{1}{3}$ of the total length) downstream from entering the thermalization tube. Therefore in the following we need to address only the exchange of heat between the metal nanoparticles in the beam and the surrounding helium bath at $T_w$. This requires knowledge of helium pressure, $P$, in the tube and the transit time of nanoparticles through the tube.

$P$ as a function of helium volume flow rate $Q_0$ set by the controller was measured by attaching a pressure gauge to the side of a dummy thermalization tube with the same dimensions as the actual one, see Fig. 4. The use of a diaphragm differential pressure gauge (Dwyer Magnehelic model 2001) ensured a direct readout unaffected by gas temperature.

The nanoparticles' transit time $t$ inside the thermalization tube can be estimated in two ways. One is to use the flow time of helium gas (this neglects the velocity slip effect[13,14] between the light gas and the heavy nanoparticles). Conservation of mass implies that the helium volume flow rate through the tube is $Q = Q_0(P_0/P)(T_w/T_0)$, where $P_0$ and $T_0$ are the "standard" 1 bar and 273 K pressure and temperature which are referenced by the sccm rate $Q_0$ displayed by the flow controller. The gas transit time through the thermalization tube is then equal to the volume $\pi(d/2)^2 l$ of the latter divided by $Q$, yielding an estimate for the nanoparticle tube transit time of

$$t \sim \frac{\pi d^2 l}{4Q_0} \frac{T_0}{T_w} \frac{P}{P_0} \tag{1}$$

For the operational helium inlet flow rate $Q_0 = 240$ sccm and the tube temperatures shown in Fig. 4, the resulting $t$ is $\sim 2$ ms.

Another estimate is obtained from a measurement of nanoparticle speeds in the outgoing beam. A rotating wheel chopper was installed downstream from the source and a multichannel scaler was used to measure the time delay between the chopper's optical switch trigger and the arrival of particles at the ion detector. For Li particles, the beam speed was found to vary from 175 m/s at $T_w = 170$ K to 195 m/s at $T_w = 370$ K.[10]

Dividing the length of the thermalization tube by these speeds we obtain $t \sim 0.5$ ms. Like in the preceding paragraph, this transit time is an underestimate, because it neglects the fact that the nanoparticle beam undergoes some aerodynamic acceleration inside and upon exit from the tube. For Na and K nanoparticles the source exit velocities are lower, reflecting a greater velocity slip due to their higher masses and implying a longer $t$.

A slow moving nanoparticle of radius $R$, experiences $\zeta = \sigma n_{He} \bar{v}$ collisions with helium atoms per unit time, where $\sigma = \pi R^2$ is the particle cross section, and $n_{He} = P/(k_B T_w)$ and $\bar{v}$ are the number density and mean thermal speed of the helium atoms, so that

$$\zeta = \pi R^2 n_{He} \sqrt{\frac{8 k_B T_w}{\pi m_{He}}} . \tag{2}$$

A 10 000-atom lithium particle ($R = 3.5$ nm, see Sec. II.E) experiences $\zeta \approx 10^9$ s$^{-1}$ for the present operating parameters. This translates into $\square 10^6$ collisions over the transit time $t$ estimated above, which appears plentiful for thermalization.[15] This assessment can be quantified as follows.

Since the mean free path of helium atoms in the tube is $\approx 300$ nm $>> R$, the energy exchange between the nanoparticle at temperature $T$ and the surrounding gas at temperature $T_w$ can be treated within the free-molecular-flow regime.[16] The heat transfer rate is represented as

$$\dot{q} = a\zeta \cdot 2k_B(T_w - T) . \tag{3}$$

Here $2k_B T_w$ represents the average kinetic energy of a helium atom striking the nanoparticle surface (a correction for the case of a nanoparticle moving with a finite speed through the gas has been derived,[17] but is not important here), and $2k_B T$ is the kinetic energy of an atom evaporating from the surface. The factor $a$ is known as the thermal accommodation coefficient[18-20] It has been measured for helium impacting sodium and potassium surfaces[19] and has a magnitude of approximately 0.1–0.2. For lithium-helium collisions there are no published data, but using $a \sim 0.1$ should be safe for a lower-limit estimate of the heat transfer rate.

Approximating the particle's heat capacity as a constant and defining $\Delta T = T_w - T$, we can rewrite Eq. (3) for an $N$-atom particle as

$$\frac{d(\Delta T)}{dt} = -\frac{a\varsigma}{N}\frac{2k_B}{c_{at}}\Delta T ,$$ (4)

where $c_{at}$ is the metal heat capacity per atom. Hence the nanoparticle temperature approaches that of the thermalization tube exponentially, with a time constant

$$\tau = \frac{N}{a\varsigma}\frac{c_{at}}{2k_B}$$ (5)

To estimate the longest $\tau$ encountered in the measurements, we can take the Dulong-Petit limit $c_{at} = 3k_B$ reached at the top of the studied temperature range, and a nanoparticle size of $N \approx 10\,000$ observed for Li, as used in the evaluation of the collision frequency $\varsigma$ above. This size is in the high range of the nanoparticle distribution (see the next section), and therefore corresponds to a worse-case estimate: the time constant for achieving thermalization scales as $N$ whereas the gas collision frequency only scales as $N^{2/3}$.

Taking $a \sim 0.1$, we find $\tau \sim 0.1$ ms. This is at least a factor of 5 shorter than the time $t$ spent by the nanoparticles inside the thermalization tube, hence their adequate equilibration with the tube walls is assured.

### D. Nanoparticle sizes

The distribution of nanoparticle sizes in the beam is measured by a homebuilt linear time-of flight (TOF) mass spectrometer of the Wiley-McLaren geometry. The particles are ionized 50 cm downstream from the skimmer using 5 ns laser pulses of 355 nm wavelength (Nd:YAG Continuum Minilite II with the third harmonic crystal) at 15 Hz repetition rate. After a 2 m free flight, the ion pulses are registered by the ion detector, sorted into time bins by the aforementioned multichannel scaler (ORTEC), and converted from the scale of arrival times to that of masses or of the number of atoms, $N$.

An example of a TOF mass spectrum is shown in Fig. 5. The leftmost peaks at low sizes are substantially affected by the laser fluence, indicating that they originate from laser-induced fragmentation. They do not contribute to the photoelectron yield measured from neutral particles downstream, hence they are disregarded in the analysis. (Additionally, even if there were small clusters present, their ionization would onset at significantly higher energies than for the main nanoparticle distribution because of the finite-size shift described below.) As a result, the near-threshold photoionization fit derives from the larger particles.

Given these considerations, we fit the data to a lognormal distribution, as shown in the figure, and use the portion of the distribution enclosed by the fit for determining the WFs. The typical mean sizes $N$ for Li, Na, and K nanoparticles are approximately 7500, 5000, and 4500 atoms, respectively, with full-width-at-half-maximum values on the order of the mean, i.e., $\Delta N \simeq N$. As detailed below, the sizes are checked before and after each ionization yield scan to ensure that the distribution did not drift during the measurement. The value of $N$ typically varied by no more than 10-15% between repeated measurements on different days and between measurements at different thermalization tube temperatures.

## E. Population of multiple charged nanoparticle ions in the mass spectra

In addition to the presence of laser-induced fragmentation, we must be conscious of another important effect. The metal nanoparticles possess substantial photoabsorption cross sections and therefore may be able to undergo sequential ionization during their exposure to the laser pulse. In that case the TOF mass spectra would contain multiply-ionized particles, and the original neutral distribution would be located at much higher sizes (e.g., $2N$ for doubly-ionized particles, $3N$ for triply-ionized, etc.) To address this possibility we need to consider the photoionization efficiency and the dependence of IE on nanoparticle size. The latter can be expressed as

$$IE_Z = WF + (\alpha + Z)\frac{e^2}{R}. \tag{6}$$

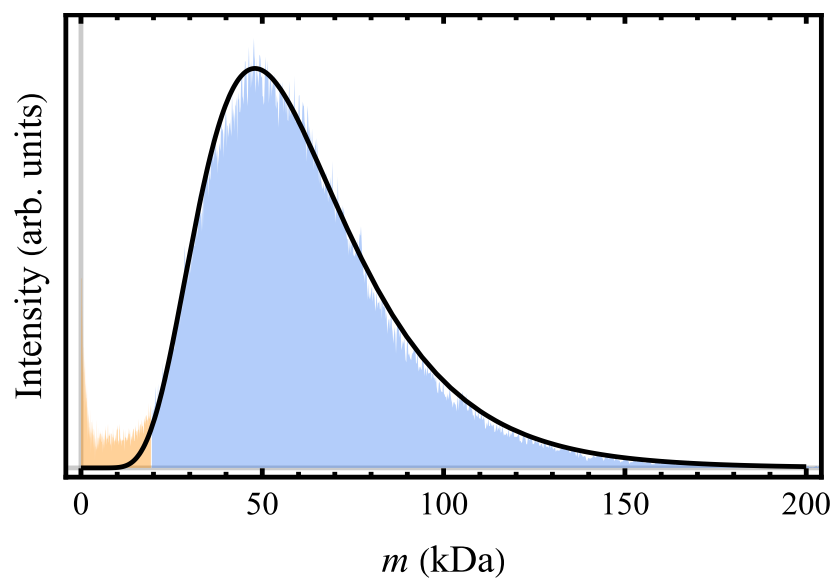

FIG. 5. An example of the mass distribution of lithium nanoparticles. The appearance and intensity of the orange portion of the TOF mass spectrum are strongly impacted by laser power, therefore this portion is assigned to laser-induced fragmentation and is not included in the analysis. The main peak does not change as laser pulse energy is varied from 0.06 mJ to 1 mJ, and therefore reflects the actual beam population, as discussed further in Sec. II.E. The peak is fitted to a lognormal distribution shown by the black solid curve, and this portion of the mass spectrum is used for analyzing the nanoparticle ionization energies.

Here $IE_Z$ is the ionization energy of a nanoparticle of initial charge $Ze$, $e$ is the magnitude of the electron charge, and $R = r_s a_0 N^{1/3}$ is the particle radius; $a_0$ is the Bohr radius, $r_s$ is the Wigner-Seitz electron density parameter (3.25 for Li, 3.94 for Na, 2.7 for K). It has been found that for alkali nanoclusters $\alpha$ is close to the value of 3/8 predicted by an image-charge argument:[21] $\alpha \approx$ 0.31-0.33 for Li,[22] $\alpha \approx 0.39$ for Na,[23] and $\alpha \approx 0.34$-0.38 for K.[24,25]

Thus as long as the photon energy exceeds $IE_Z$, the production of cations of charge $(Z+1)e$ over the duration of the laser pulse is in principle possible. The probability of multiple ionization grows with increasing photon flux, i.e., with increasing intensity of the laser pulses.

Consider the sequential photoionization of nanoparticles irradiated by a laser pulse of energy $E_p$. The average number of ionizing photons absorbed by the particle is given by $n_0 = \sigma_i E_p / (E_\gamma A)$, where $\sigma_i$ is the ionization cross section, $E_\gamma$ is the photon energy, and $A$ is the effective laser spot area. Single-photon ionization can occur only if the photon energy exceeds the ionization threshold, which increases with each ionization step according to Eq. (6). Thus the maximum achievable charge state $Z_{max}$ is that for which $IE_{Z_{max}-1} \leq E_\gamma < IE_{Z_{max}}$. The probability of reaching a charge state $Z$ ($\leq Z_{max}$) follows a Poisson distribution:

$$P(Z) = \frac{n_0^Z e^{-n_0}}{Z!} \tag{7}$$

This assumes that the ionization efficiency remains the same for higher charge states. It is not a reliable approximation for small cluster sizes but should be adequate for the larger nanoparticles studied here. While the Poisson distribution mathematically allows for arbitrarily high photon absorption, here its range is constrained by $Z_{max}$.

Modeling the full size distribution of nanoparticles with multiple charge states in a laser ionization mass spectrum is challenging, since it would require knowledge of the cross-sections $\sigma_i$. Unfortunately, for large clusters and nanoparticles no comprehensive information is available regarding the magnitude of these cross sections and their size dependence. Instead, one can make use of mass spectra recorded at different laser powers to obtain an estimate of the average number of absorbed photons and to assess the charge state distribution.

For example, Fig. 6(a) shows how the mass spectrum of large sodium nanoparticles (aggregated using argon instead of helium) varies upon ionization with 355 nm laser light with different pulse energies. The particles have an average diameter of 15 nm and $IE = 2.7$ eV, enabling charge states up to $Z_{max} = 4$. The strong shift toward smaller mass-to-charge ratios for the high-power case derives from the appearance of multiply charged ions. If we assume that the low-power mass spectrum contains mostly singly-charged nanoparticles, it can be convoluted with Eq. (7) to fit the high-power mass spectrum,[26] for which we assume that it contains all possible charges up to $Z_{max}$. From this procedure we obtain an estimate of the average number of ionizing photons absorbed when the laser is operated at high power: $n_0 = 3.5$.

In contrast, the mass spectrum shown in Fig. 6(b) is much less affected by a similar variation in laser intensity. These nanoparticles have an average diameter of 8 nm, with $IE = 2.95$ eV and $Z_{max} = 2$ for 355 nm ionization. Because now both the cross section and $Z_{max}$ are lower, the average number of ionizing photons absorbed in this case is much smaller: $n_0 = 1$. As a consequence, correcting this mass spectrum for the presence of doubly ionized particles would have a minimal effect on the work function determination (described in Sec. III), resulting in a shift of at most 5 meV.

Mass calibration during actual ionization threshold measurements employed laser output powers even lower than in the preceding example. Furthermore, the laser beam was less tightly focused on the nanoparticle beam. As a result, time-of-flight mass spectra did not change as the laser pulse energy was varied, as described in the caption to Fig. 5. Thus, even allowing for the approximate nature of the presented estimates, it is safe to conclude that in these measurements multiple ionization had a negligible effect on the work function data.

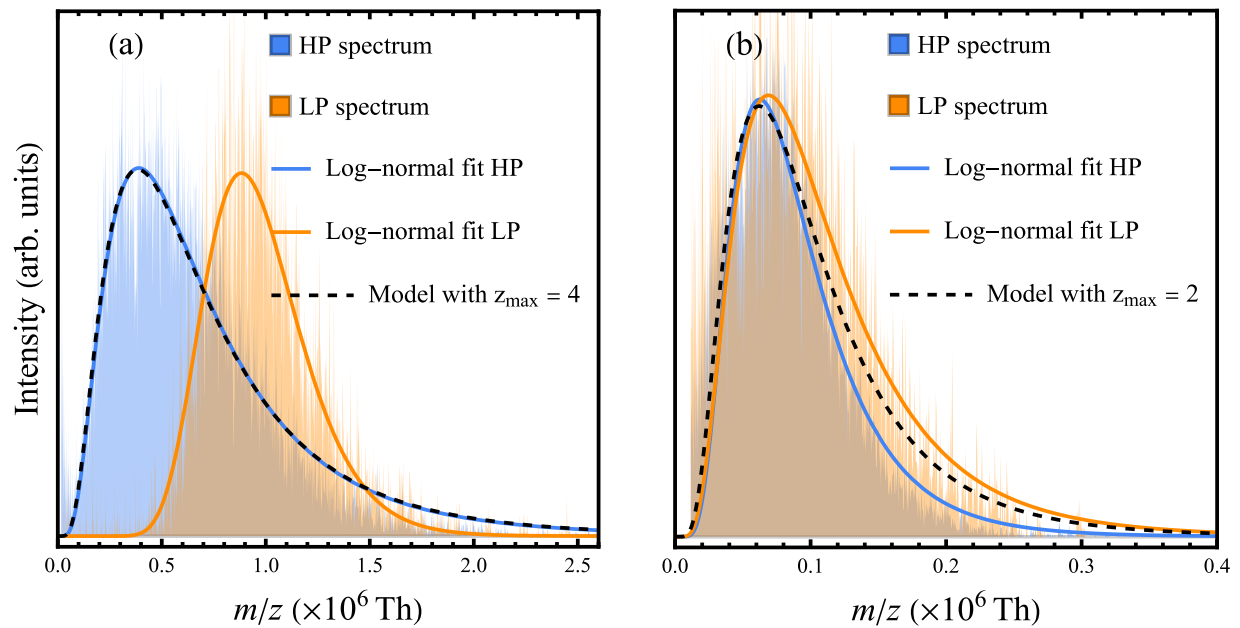

**FIG. 6.** TOF mass spectra of sodium nanoparticles generated under different gas aggregation conditions in the source: (a) Argon, $Q_0 = 240$ sccm, $T_w = 130$ K. and (b) Helium, $Q_0 = 240$ sccm, $T_w = 250$ K. Condensation in argon produces much larger particles. In both cases the nanoparticles were ionized using a 355 nm pulsed laser, and mass spectra were recorded at low (orange area, labeled LP, $E_p = 0.1$ mJ) and high (blue area, labeled HP, $E_p = 3.5$ mJ) pulse energies. The shift at higher laser intensity is due to sequential absorption of multiple photons during the ionizing laser pulse. The solid lines are fits of the data to a log-normal distribution, and the dashed lines are model fits to the Poisson distribution of ionization probabilities discussed in Sec. 2.E.

### F. Photoionization yield stage

The neutral nanoparticles are photoionized near the end of the beam flight path. Light is produced by a regulated broadband arc lamp (1000 W Hg-Xe or UV-enhanced Xe, Oriel Instruments) and travels through a distilled water IR filter, a Bausch & Lomb monochromator, and a filter for removing any second-order diffraction peaks. The monochromator bandwidth and absolute wavelength calibration were better than 1 nm and 0.4 nm, respectively, as determined using discrete lines from the arc lamp and a spectrometer (StellarNet BLUE-Wave, 0.40 nm resolution).

After passing through a rotating wheel optical chopper, it is focused into the extraction region of the ion detector. The resulting positive ion pulses are recorded in synchronization with the chopper either by the multichannel scaler or by two coordinated gated counters (National Instruments).[10] The photoion count rates range from ~5 per second in the ionization threshold region to ~$10^3$ per second farther above threshold.

In order to quantify the ionization yield it is necessary to normalize the ion counts to the nanoparticle beam intensity and to the ionizing photon flux. The irradiance at each wavelength is measured by moving a calibrated photodiode (PerkinElmer Optoelectronics VTB1112) into the ionization focal point. Its output is read via a lock-in amplifier referenced to the optical chopper. The intensity of the incoming nanoparticle beam is monitored by focusing light from a reference LED (365 nm, Violumas VC2X2C45L9-365 stabilized by a constant-current source) into the same point.

After the beam source is heated and stabilized, data acquisition proceeds as follows.

(i) A TOF mass spectrum is acquired.

(ii) The photodiode is positioned in the ionization region and records the irradiance as the monochromator is tuned through the set of selected wavelengths.

(iii) The UV LED is positioned in from of the focusing lens and the nanoparticle beam intensity is measured.

(iv) The monochromator is stepped through the selected wavelengths and the corresponding photoion signal is recorded by the gated counters.

(v) The LED is used once more to ensure that nanoparticle beam intensity has not changed by more than 10% during the yield curve measurement.

(vi) Finally, another TOF mass spectrum is acquired to ensure that the particles distribution has remained stable.

The monochromator dial, the LED mount, and the photodiode's linear motion feedthrough are controlled by stepper motors operated by Arduino Uno boards. The data acquisition sequence (ii-v) is fully automated via a LabVIEW program interfaced with the Arduino boards via serial communication and with the counters on a National Instruments data acquisition board.

This setup has proven to be reliable and efficient, allowing us to collect stable yield curves at a rate of approximately ten per hour. Fig. 7 illustrates such a sequence, showing excellent reproducibility between data sets. To further demonstrate the stability of the data, the figure shows that the IEs between data sets fluctuate by less than 1%, and the resulting average IE value has a standard deviation of less than 0.2%.

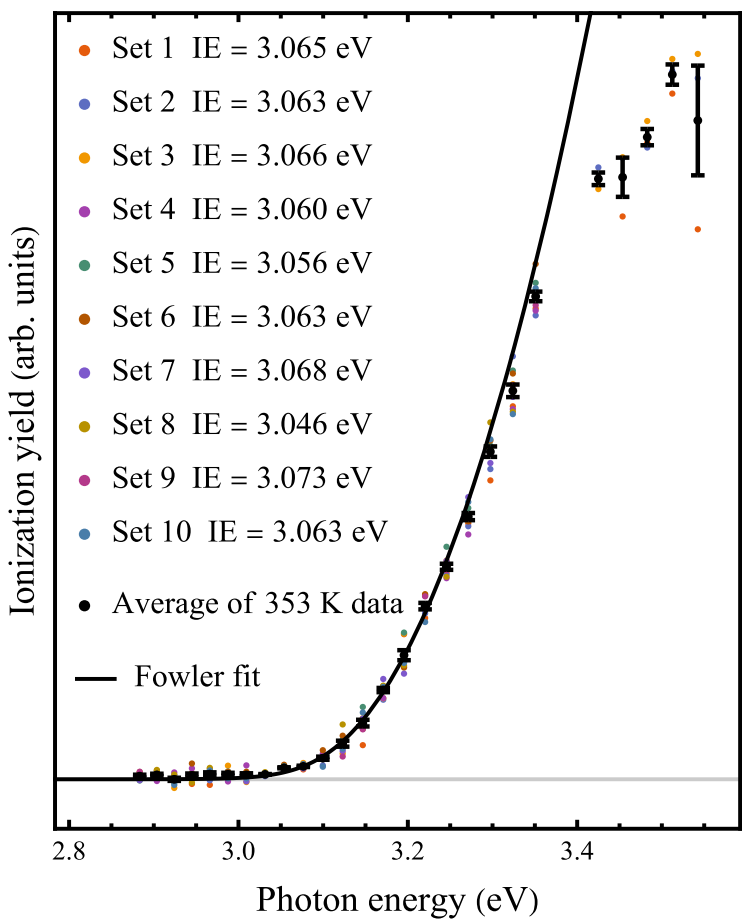

**FIG. 7**. Photoionization yield data for Li nanoparticles of average size $N = 7100$ atoms at $T = 353$ K thermalization tube temperature. The points are normalized yields at different photon energies, with a wavelength step of 3 nm. Colors represent separate data sets, each one taking approximately 6 minutes to acquire. The black dots and their error bars are the average and standard deviation for each wavelength. The solid line is a Fowler fit (Sec. III) to the black dots, and the legend tabulates the IEs determined by the Fowler fit for each individual data set. The average value is 3.064 eV and the standard deviation is 0.007 eV. (Above 3.4 eV the points begin to deviate from the curve and are not included in the fit, see Sec. III.)

## III. DATA ANALYSIS

As can be seen from Fig. 7, for an accurate determination of IEs it is very important to employ a consistent treatment of the threshold region. As was found for coinage metal aerosol particles[27] and then corroborated for metal nanoclusters of various sizes and materials,[6,25,28-30] the Fowler law,[2,31,32] originally formulated for bulk metallic surfaces, provides an excellent, tractable, and elegantly uniform fit to frequency- and temperature-dependent photoionization profiles.

The formula is obtained by evaluating the flux of those conduction electrons whose kinetic energy of motion perpendicular to the surface, plus the energy $h\nu$ contributed by an absorbed photon, exceeds the WF. At finite temperatures, when thermal smearing of the Fermi–Dirac distribution is accounted for, it is found that the photoelectron yield $Y$ satisfies

$$\log\left(\frac{Y}{T^2}\right) = B + \log f\left(\frac{h\nu - IE}{k_B T}\right), \tag{8}$$

where we replaced WF with IE in the argument on the right-hand side. $B$ is a temperature-independent coefficient incorporating constants and instrumental parameters, and $f$ is a certain integral over the distribution function. The function $f$ is expressible in terms of Jonquière's polylogarithm $Li_n(x)$ with $n = 2$ and $x = -\exp[(h\nu-IE)/k_B T]$. A plot of $[\log(Y/T^2) - B]$ vs. $(h\nu - IE)/k_B T$ is known as a "Fowler plot," and via a two-parameter fit of the data to the universal curve $\log f(x)$ one obtains the photoemission threshold.

We have developed an procedure to analyze ionization yield plots such as that shown in Fig. 7. First of all, it is necessary to establish the frequency range which will be fitted to the Fowler curve. As can be seen from the figure, above a certain point the yield begins to deviate from the theory curve. This occurs both because additional interaction channels, for example surface plasmon resonances, become available,[5] and because the photon energy begins to exceed the region where Fowler's derivation is applicable.[33] In addition to the upper limit for the data points to be included in the Fowler fit, it is also necessary to identify the lower limit, because including excessively many sub-threshold points tends to induce noise and enlarge the error bars.

We determine the usable region by repeatedly fitting the Fowler curve, Eq. (8), to the averaged photoemission data (such as the black points in Fig. 7) and using each fit's threshold to decide which points to keep for the next iteration. We start by fitting the full dataset to obtain an initial threshold estimate. We then discard all data points below that threshold, so that only the rising-edge portion remains, and refit the curve to this dataset. Because this new threshold will generally shift slightly, we again remove any points below it and refit. After several such fit-remove-refit cycles, the threshold converges to the best self-consistent starting point. The upper limit for the data set is then identified by adjusting the endpoint to obtain a configuration with the least residual sum of squares for the maximal number of data points. We found that this iterative algorithm yields a robust selection of the fitting interval satisfying the Fowler plot condition. Fig 8 illustrates the resulting universal behavior fitted by the described algorithm.

Having identified the fitting interval based on the averaged data for a given temperature, we proceed to determine the nanoparticle IE from individual data sets, and then extrapolate it to the bulk limit to deduce the bulk WF for this $T$. There are two options for the latter procedure.

The simpler option is to obtain the IE from a collection of data sets and then to subtract the average size correction for this collection run. Here the best-fit IE and its uncertainty are obtained by a Monte Carlo resampling procedure. This is done by randomly selecting a data point at each given wavelength (that is, by randomly picking one of the colored dots in Fig. 7 for each photon energy in the fitting interval) and fitting the resulting Fowler plot. This is repeated for 2000 different random combinations, resulting in a value for the IE that remains unchanged to within 0.1 meV if the resampling procedure is rerun.

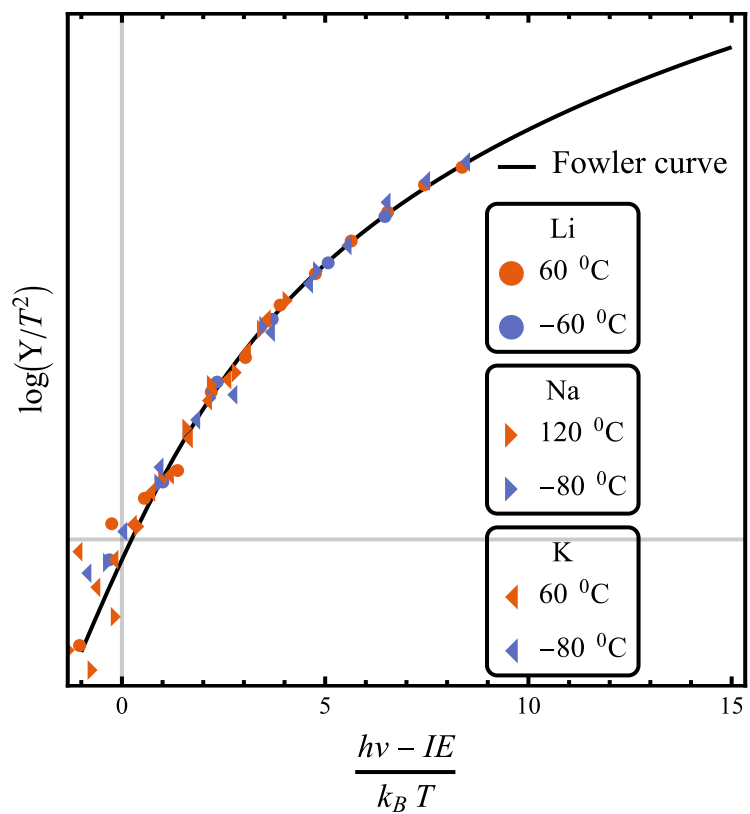

**FIG. 8**. The photoionization yields for nanoparticles of different metals and temperatures align along the universal Fowler plot. Each data point is the average of several different runs.

The WF is then obtained by using Eq. (6) with Z=0 and the average nanoparticle radius. The latter is determined from the TOF mass spectrum,[34] averaged over steps (i) and (vi) in the data acquisition cycle described in Sec. II.E:

$$WF = IE - \frac{\alpha e^2}{r_s a_0} \left\langle \frac{1}{N^{1/3}} \right\rangle \tag{9}$$

A more computationally expensive method is to represent the experimentally measured yield as being formed by individual contributions from all the different cluster sizes $N$ in the TOF mass spectrum:

$$Y = \sum_N Y(N) I(N),$$

$$Y(N) = e^B T^2 f \left( \frac{h\nu - WF + \alpha e^2 \left( r_s a_0 N^{1/3} \right)^{-1}}{k_B T} \right), \tag{10}$$

where $I(N)$ is the abundance of each size $N$ in the mass spectrum. We average 100 randomly selected points from each run and fit 100 such sets to Eq. (10). (*Mathematica*'s built in `parallelize` function is used to accelerate the computation.) This produces roughly the same tolerance as the combination of Eqs. (8) and (9) repeated 2000 times, as described above.

The average difference between the two fitting methods for all temperatures is approximately 7 meV, and the results shown below are the averages of the two. Furthermore, for both methods the parameter $B$ remains stable within individual measurement runs and varies by no more than 8% over the full temperature range (due to small shifts in the beam alignment and in the photoionization cross section), with no effect on the precision of the ionization threshold fit. This high degree of stability affirms the physical meaning of the fit and its quality.

We also incorporated the effect of the finite bandwidth of the monochromator, described in Sec. II.F, by adding random sampling of $h\nu$ from a 1 nm-wide distribution around its nominal settings. This added a 3 meV variation to the fitted $WF$ values.

## IV. WORK FUNCTIONS

Table 1 lists the results obtained for K, Na, and Li at specified thermalization tube settings. The latter are determined by the fact that at higher temperatures these nanoparticles undergo a melting transition.[35,36] Each measurement consisted of ten data sets, as in Fig. 7, and was repeated three times on different days. The error of the mean is calculated in quadrature from variations between fitted results from each data set, between the published coefficients $\alpha$ [Eq. (6)], between the outputs of the two fitting methods described in the preceding section, and those due to the spectral resolution of the ionizing light. The other values in the table are those quoted in the literature for polycrystalline samples (sample temperatures are usually not specified). The results are in good agreement, validating the technique presented here.

**Table 1**: Alkali metal work functions, in eV.

|  | Present work | References 37,38 |
|---|---|---|
| K | 2.329±0.005 ($T_w$ = -40 °C) | 2.30, 2.29±0.02 |
| Na | 2.762±0.005 ($T_w$ = 0 °C) | 2.75, 2.54±0.03 |
| Li | 2.953±0.007 ($T_w$ = 20 °C) | 2.9, 2.90±0.03 |

## V. CONCLUSIONS AND OUTLOOK

The study of isolated nanoparticles has value not only in the discovery of novel effects on the nanoscale, but also in providing access to ultrapure surfaces for benchmarking and modeling important bulk processes. In this paper we described a nanoparticle beam setup which is capable of highly efficient, automated, and precise determination of metal work functions.

The method was successfully demonstrated for alkali metals, whose surfaces are normally challenging to probe due to high reactivity. The achieved precision of a fraction of a percent over a wide temperature range will enable accurate studies of the influence of temperature and vibrational dynamics on the work functions. For example, WF shifts due to melting or to isotopic substitutions can be resolved.[10,36]

The described setup employed a vapor condensation nanoparticle beam source and an arc lamp. By using other nanoparticle[7,39] and radiation sources, and introducing a higher degree of mass selectivity (for example, by extracting the photoionization products into a Wien filter[40] or a high-mass-range reflectron time-of-flight mass spectrometer[41]) it should be possible to achieve comparable precision for other metals, potentially approaching the record sub-meV level recently attained for gold surfaces using angle-resolved photoemission spectroscopy,[42] but for more reactive surfaces, without a need for ultrahigh vacuum, and over a wide range of temperatures.

Furthermore, since contemporary nanoparticle sources make it possible to generate a variety of alloyed and core-shell structures, interesting insights into their properties may be derived from accurate temperature-dependent photoionization measurements.

## ACKNOWLEDGMENTS

We would like to thank Prof. Bernd von Issendorff for valuable advice. We also thank Caleb Blumenfeld, Derrick Korponay, and Rayan Mendiola for capable assistance with the experiments, Stephanie Silva for help with graphics, and the staff of the USC Machine Shop for exceptional technical help and skillful machining. S.P. acknowledges financial support by the Vienna Doctoral School in Physics. This research was supported by the U.S. National Science Foundation under Grant No. DMR-2003469.

## AUTHOR DECLARATIONS

### Conflict of Interest

The authors have no conflicts to disclose.

### Author Contributions

A. A. Sheekhoon: Conceptualization (equal); Data curation (equal); Formal analysis (equal); Investigation (equal); Methodology (equal); Software (equal); Visualization (equal); Writing/original draft preparation (equal). A. O. Haridy: Conceptualization (equal); Data curation (equal); Formal analysis (equal); Investigation (equal); Methodology (equal); Software (equal); Visualization (equal); Writing/original draft preparation (equal). S. Pedalino: Conceptualization (supporting); Formal analysis (supporting); Investigation (supporting); Methodology (supporting); Writing/original draft preparation (supporting). V. V. Kresin: Conceptualization (lead); Formal analysis (supporting); Funding acquisition (lead); Methodology (supporting); Project Administration (lead); Writing – original draft preparation (equal).

## DATA AVAILABILITY

The data that support the findings of this study are available from the corresponding author upon reasonable request.